\documentclass[aps,prb,reprint,superscriptaddress]{revtex4-1}
\usepackage{H1}
\usepackage{amsmath}
\usepackage[pdftex,colorlinks=true]{hyperref}
\hypersetup{
citecolor = blue
}
\usepackage[normalem]{ulem}
\usepackage{amssymb}
\usepackage{amsthm}
\usepackage{amsfonts}
\usepackage{bbm}
\usepackage{array}

\def\bra#1{\mathinner{\langle{#1}|}}
\def\ket#1{\mathinner{|{#1}\rangle}}

\begin{document}
\title{A Floquet Model for the Many-Body Localization Transition}

\author{Liangsheng Zhang}
\affiliation{\mbox{Department of Physics, Princeton University, Princeton, NJ 08544, USA}}

\author{Vedika Khemani}
\affiliation{\mbox{Department of Physics, Princeton University, Princeton, NJ 08544, USA}}
\affiliation{\mbox{Department of Physics, Harvard University, Cambridge, MA 02138, USA}}

\author{David A. Huse}
\affiliation{\mbox{Department of Physics, Princeton University, Princeton, NJ 08544, USA}}
\affiliation{\mbox{Institute for Advanced Study, Princeton, NJ 08540, USA}}

\begin{abstract}
The nature of the dynamical quantum phase transition between the many-body localized (MBL) phase and the thermal phase remains an open question, and one line of attack on this problem is to explore this transition numerically in finite-size systems.  To maximize the contrast between the MBL phase and the thermal phase in such finite-size systems, we argue one should choose a Floquet model with no local conservation laws and rapid thermalization to ``infinite temperature'' in the thermal phase.  Here we introduce and explore such a Floquet spin chain model, and show that standard diagnostics of the MBL-to-thermal transition behave well in this model even at modest sizes. We also introduce a physically motivated spacetime correlation function which peaks at the transition in the Floquet model, but is strongly affected by conservation laws in Hamiltonian models. 
\end{abstract}

\maketitle

\section{Introduction}
Many-body localization (MBL) generalizes the phenomenon of Anderson localization to the interacting setting\cite{Anderson58, Basko06, PalHuse, OganesyanHuse, Nandkishore14, AltmanVosk}. Many-body localized systems do not reach local thermal equilibrium, and even highly excited states retain local memory of their initial conditions to arbitrarily late times. The transition between the MBL and thermalizing phases is not a conventional thermodynamic phase transition and lies outside the framework of equilibrium statistical mechanics and critical scaling theory. Instead, this transition is a novel dynamical phase transition which involves the restoration of statistical mechanics at the level of individual eigenstates --- across this {\it eigenstate phase transition} \cite{Huse13, PekkerHilbertGlass, Bauer13,Vosk14, Kjall14, Chandran14, Bahri15, Khemani15} the nature of the system's many-body eigenstates changes in a singular fashion from thermally and ``volume-law'' entangled eigenstates in the thermal phase that obey the eigenstate thermalization hypothesis (ETH) \cite{Deutsch, Srednicki, Rigol}, to nonthermal and only boundary-law entangled eigenstates in the MBL phase\cite{PalHuse, Bauer13}.

While the MBL-to-ETH transition has attracted much recent interest \cite{Basko06, OganesyanHuse, PalHuse, Kjall14, BarLevDynamics,VHA,PVP, ZhangRG, SerbynCriterion,AgarwalGriffiths2015,Luitz15,santos,Gopalakrishnan15,Devakul15,goold,mobilityedge, SerbynSpectralStats,CLO,GopalakrishnanGriff, KhemaniTransition}, very little is definitively understood about its properties. Much of the work studying this transition has involved numerical exact diagonalization (ED) of one dimensional spin-chains for small system sizes $L \leq 22$.  All of these ED studies have found apparent critical exponents that violate Harris/Chayes type bounds\cite{Harris, CCFS1, CCFS2, CLO}, and have observed a finite-size crossover only on the thermal side of the phase transition. On the other hand, approximate renormalization group studies\cite{VHA, PVP, ZhangRG} find a continuous transition with exponents that obey the Harris/Chayes inequality. A recent study\cite{KhemaniTransition} has found that aspects of this transition even look discontinuous, in that quantities like the entanglement entropy of small subsystems appear to vary discontinuously across the transition. All in all, it is safe to say that we have not (yet?) found methods that allow a detailed and convincing study of the properties of this transition. Since most approaches---particularly numerical studies using microscopic spin chains---have considerable freedom in choosing specific models for studying the transition, here we address the question of what features one might possibly look for in choosing the ``best'' type of model to study.

The many-body localized phase is, in at least some cases, now understood to be a new type of integrable system with an extensive number of localized conserved quantities \cite{Serbyn13cons,Huse14}.  The phase transition into the MBL phase from the thermal side can thus be thought of as the emergence of all these new conservation laws.  This suggests that an important property of the thermal phase may be its lack of localized conserved quantities.  In finite sized systems any conserved quantity is, in some sense, localized due to the finite size. Thus, one can argue that the best representative of the thermal phase might be a phase with no local conservation laws, not even energy.  Such models are available in the form of so-called Floquet models with time-periodically driven Hamiltonians $H(t) = H(t+T)$. 

Generically, interacting Floquet models have no local conservation laws or symmetries. When they thermalize, they maximize the entropy without any constraint and so, in a certain sense, they thermalize to infinite temperature\cite{Rigol14, LazaridesHeating, Ponte15b, KIH}. Having no conservation laws also allows Floquet models to thermalize faster and more completely than the corresponding Hamiltonian models that have conserved energy slowing down their thermalization \cite{LHD}. It has recently been shown that with sufficiently strong disorder and weak enough driving, Floquet systems can also be many-body localized\cite{Lazarides14, Ponte15, Ponte15b, Abanin16} and exhibit a dynamical phase transition to the ``infinite-temperature'' thermal phase as a function of the driving and/or of the randomness.  The properties of the Floquet system are determined by the eigenstates of the Floquet unitary, $U = \mathcal{T} e^{-i\int_{0}^T dt H(t)}$, which is the time-evolution operator for one period. In the MBL phase, the Floquet eigenstates have boundary-law entanglement and can display novel forms of quantum orders, some of which are unique to the driven setting\cite{Khemani15}. Recent work has identified and classified several Floquet phases with symmetry-breaking and topological order\cite{Khemani15, vonKeyserlingkSondhi16a, vonKeyserlingkSondhi16b, Else16, Potter16, RoyFloqSPT, CVS}.

The strongly thermalizing features of the Floquet-ETH phase make such models potentially good candidates for studying the MBL transition. In this paper, we introduce a specific such spin-1/2-chain Floquet model with no conservation laws.  In the limit of zero disorder this model becomes the model introduced in Ref. \onlinecite{KIH} as an example of a Floquet model that thermalizes very rapidly \cite{LHD,KimSlowestLocal}.
We show results for several diagnostics which are frequently used to characterize the MBL phase transition. These demonstrate that the model can indeed reach a fully thermalized phase as well as a fully localized phase, and they show that the standard indicators of the phase transition behave well for this model already at modest system sizes. We also study end-to-end spacetime correlation functions which probe physical, experimentally measurable features of the MBL phase and the transition.  
Correlation functions play an important role in studying ``traditional" phase transitions, but appear to be under-represented in numerical studies of MBL phase transitions to date, most likely because these are strongly influenced (at small sizes) by the presence of even a few conservation laws like energy and total $\sigma^z$.  By using a Floquet model with no conserved densities, we remove the effects of these conservation laws on the long-distance spin correlations, making them more promising tools for investigating the MBL phase transition.

\begin{figure*}[ht]
\includegraphics[width=2\columnwidth]{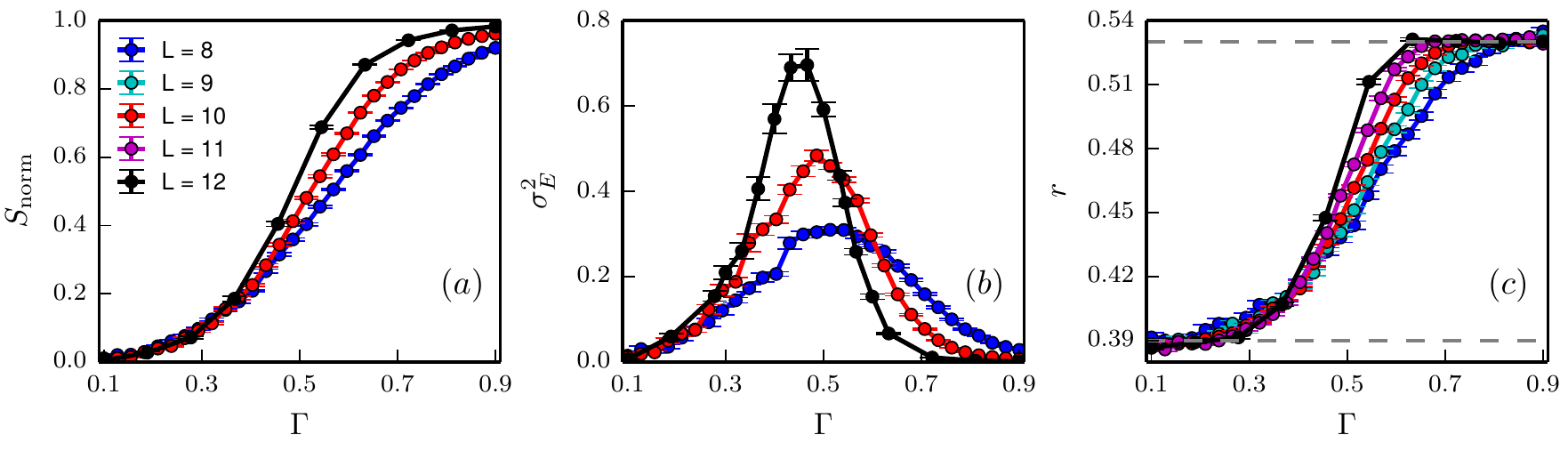}
\caption{(color online).  (a) The normalized average entanglement entropy, chosen so a random pure state has $S_{\rm norm}=1$.   Our estimate of the critical
point from other data, shown below, is $\Gamma_c\cong 0.3$, with the MBL phase at $\Gamma < \Gamma_c$ and the thermal phase at $\Gamma > \Gamma_c$. (b) The variance $\sigma_E^2$ of the eigenstate entanglement entropy. 
In the thermodynamic limit, $\sigma_E^2$  approaches zero in the thermal phase, and apparently diverges near the phase transition. (c) The level statistics parameter $r$, as defined in the text.  In the localized phase, $r$ approaches $\cong 0.39$ for large $L$ due to Poisson level statistics, while the value approaches $\cong 0.53$ in the thermal phase due to circular orthogonal ensemble statistics \cite{Rigol14}. These two limiting values are indicated by the dashed lines.}   
\label{fig:Trans}
\end{figure*}

\section{The Model}
We study a spin-1/2 chain with nearest neighbour Ising interactions $\sigma_j^z\sigma_{j+1}^z$. To make the system nonintegrable and rapidly thermalizing deep in the thermal phase, we add both a transverse and a longitudinal field.  We choose to put the disorder only in the longitudinal field, and the Floquet unitary is given by:

\begin{equation}
U = \exp\left[-i\frac{\tau}{2}H_x\right]\;\exp\left[-i\tau H_z\right]\;\exp\left[-i\frac{\tau}{2}H_x\right],
\label{eq:model}
\end{equation}
where 
\begin{align*}
H_x &= \sum_{j=1}^{L}g\Gamma\sigma_j^x \nonumber \\
H_z &= \sum_{j=1}^{L-1}\sigma_j^z \sigma_{j+1}^z+\sum_{j=1}^{L}(h+g\sqrt{1-\Gamma^2} G_j)\sigma_j^z,
\end{align*}
and $\sigma_{i}^{\alpha = \{x,z\}}$ are Pauli spin-1/2 matrices on site $i$. We work with with open boundary conditions and fixed parameters $ (g,h,\tau) = (0.9045, 0.8090, 0.8) $, as in Refs.~\onlinecite{KIH,LHD}.
$\{G_j\}_{j=1}^L $ is a set of independent Gaussian standard normal random variables.
The model corresponds to driving the system periodically with period $ 2\tau $  such that $ H_x $ and $H_z$ each act for a time interval of length $\tau$ during a period. For convenience, we choose our reference time point at the ``center'' of $H_x$'s turn so that $ U $ has time reversal symmetry, $U^{\dagger} = U^{*}$, which makes all eigenvectors real (in the basis of product states of $\sigma^z$), thereby making numerical diagonalization of $U$ more efficient. $ \Gamma $ controls both the strength of the transverse field and the disorder strength since it scales the Gaussian random longitudinal fields.  {The form we have have chosen is such that the total mean-square field on each spin is independent of $\Gamma$.}  As $ \Gamma\to1 $, the model becomes the nonrandom Floquet model used in Refs.~\onlinecite{KIH,LHD} so it thermalizes rapidly with only weak finite size effects. As $ \Gamma\to0 $, the transverse field is turned off and the model is trivially localized with its conserved operators (its ``l-bits''\cite{Huse14}) being simply the $\{\sigma_i^z\}$. Tuning $ \Gamma $ between 0 and 1 thus gives us the dynamical phase transition between the MBL phase and the thermal phase.

\section{Diagnostics of the Transition}
In the following, we use exact diagonalization of the Floquet unitary \eqref{eq:model} to investigate several quantities near the MBL phase transition which demonstrate the ability of our model to give reasonably sharp results even for relatively small system sizes.

\subsection{Entanglement entropy}
We consider the bipartite eigenstate entanglement entropy $S_E$ given by the von Neumann entropy of the reduced density operator of a half chain: $S_E=-\mathrm{Tr}\{\rho_L \log_2\rho_L\}=-\mathrm{Tr}\{\rho_R \log_2\rho_R\}$, where $\rho_{L/R}$  are the reduced density operators of the left/right half chains when the full chain is in an eigenstate of $U$.
When $ \Gamma\to1 $, Floquet dynamics thermalizes these half chains to infinite temperature \cite{Rigol14, LazaridesHeating, Ponte15b, KIH} so, for large enough
$L$, all eigenstates have ``volume law'' entanglement entropy close to the Page value\cite{Page} for random pure states \cite{LHD}:  
\begin{equation}
S_R(L)=\frac{L}{2}-\frac{1}{2\ln 2}-\mathcal{O}\left(\frac{1}{2^L}\right) ~.
\label{eq:page}
\end{equation}
On the other hand, in the localized phase, eigenstates of $U$ have ``boundary-law'' entanglement entropy\cite{Nandkishore14, Bauer13}, so 
$ S_E / S_R\to 0 $ with increasing $L$ in the MBL phase.
Fig.~\ref{fig:Trans}(a) shows the normalized mean entanglement entropy $S_ {\rm norm} \equiv  [\overline{S_E}] / S_R $ across the transition as $\Gamma$ and $L$ are varied.
Here $ \overline{\cdots} $ indicates averaging over all eigenstates for one disorder realization, and $ [\cdots] $ indicates averaging over disorder realizations.  
This normalized entropy transitions from 0 to 1 as $ \Gamma $ increases, with the transition becoming a steeper function of $\Gamma$ as $L$ is increased, as expected and as also seen in Hamiltonian models \cite{Kjall14,Luitz15}.

A sensitive test of the MBL-to-ETH transition\cite{Kjall14} is the variance of the distribution of $S_E$ defined as 
$\sigma^2_E = [\overline{ (S_E-S_{\rm ave})^2 }] $, where $ S_{\rm ave} $ is the average entanglement entropy over
all eigenstates from all realizations of the disorder. 
The data in Fig.~\ref{fig:Trans}(b) shows a peak in this quantity which increases strongly with $L$ and occurs on the thermal
side of the phase transition; the location of the peak approaches the transition as $L$ is increased. 
This entanglement variance is expected to vanish in the limit of large $L$ in the thermal phase (as entropies of all eigenstates approach $ S_R $) and to remain boundary law in the localized phase. 
These expected trends are clear in Fig.~\ref{fig:Trans}(b), even though the lengths sampled are modest.

\subsection{Level statistics}
Level statistics are a convenient measure of the level repulsion in a system and are used to distinguish localized and thermal phases.
For a Floquet system, the eigenvalues of $U$ are unimodular and can be denoted as $e^{i\theta_n}$, with the $2^L$ eigenvalues labeled by integers $n$ consecutively around the unit circle.  The level statistics ratio is defined as 
 \cite{Ponte15b,OganesyanHuse,PalHuse,atas} $ r=[\overline{\min{(\Delta\theta_n, \Delta\theta_{n+1})}/\max{(\Delta\theta_n, \Delta\theta_{n+1})}}] $,
 where $ \Delta\theta_n = \theta_n - \theta_{n-1} $.
In the localized phase, $ r\cong 0.39 $ at large $L$ following Poisson level statistics, while in the thermal phase $r\cong 0.53$ following circular orthogonal ensemble level statistics \cite{Rigol14}.
Fig.~\ref{fig:Trans}(c) shows that $r$ becomes a steeper function of $\Gamma$ as $L$ is increased,
and the system is well thermalized/localized at the two ends of the range of $ \Gamma $ even for a size as small as $L=10$.
We note that our model has an inversion symmetry about the center of the chain at $ \Gamma=1 $ (no disorder), which would lead to accidental near-degeneracies between even and odd parity sectors, disguising the level repulsion of thermal states.
However, this is not a problem here as Fig.~\ref{fig:Trans}(c) shows that the system is well thermalized by $ \Gamma\simeq 0.9 $ and still has enough disorder to guard against this inversion symmetry effect.

\subsection{Magnetization Imbalance}
\begin{figure}[h]
\includegraphics[width=\columnwidth]{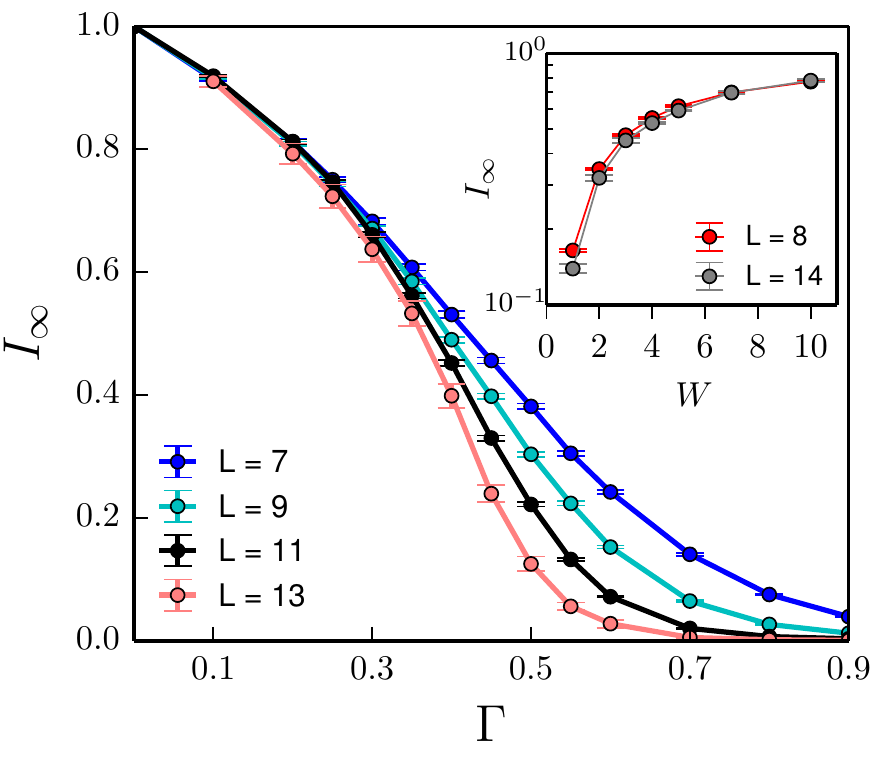}
\caption{(color online). The late time-averaged magnetization imbalance $I_\infty$ as defined in \eqref{eq:imbalance} plotted for the Floquet model. In the thermal phase (large $\Gamma$), $I_\infty$ approaches 0 for large $L$ as the system thermalizes, while $I_\infty$ remains non-zero in the localized phase since some local memory of the initial magnetization patterns persists for infinitely long times. (inset) Same quantity plotted for a Hamiltonian model shows that the presence of conservation laws impedes thermalization in the thermal phase (small $W$) and leads to a comparatively large, non-zero $I_\infty$ even deep in the thermal phase. }
	\label{fig:imbalance}
\end{figure}

Next, we consider a dynamical diagnostic which directly probes the breakdown of thermalization in the MBL phase. Following the experimental setup in Ref.~\onlinecite{Schreiber2015}, the system is initially prepared in a state with staggered magnetization $|\psi_0\rangle = |\uparrow \downarrow \uparrow \cdots \uparrow\downarrow\rangle$. The relaxation of this initial state is quantified via the ``imbalance'' in magnetization between the even and odd sublattices defined as $$I(t) = \left \langle \psi(t) \left|\frac{\sum_{i \in \rm even} \sigma_i^z - \sum_{i \in \rm odd} \sigma_i^z}{L}\right| \psi(t)\right \rangle,$$
and $I(t=0) = 1$. While $I(t)$ quickly relaxes to zero in the thermal phase, it remains non-zero even at infinitely late times in the localized phase since some memory of local initial conditions persists forever\cite{PalHuse, Nandkishore14,VasseurRevivals}. We define the infinite time-averaged and disorder averaged imbalance as 
\begin{align}
I_{\infty} &=\left[ \lim_{T \rightarrow \infty} \frac{1}{T} \int_{0}^{T} dt\;  I(t) \right] \label{eq:imbalance} \\
&= \left[\sum_n |\langle \psi_0|n\rangle|^2  \left \langle n \left|\frac{\sum_{i \in \rm even} \sigma_i^z - \sum_{i \in \rm odd} \sigma_i^z}{L}\right| n\right \rangle\right],\nonumber
\end{align}
where the sum is over Floquet eigenstates $\{|n\rangle\}$, and the second line follows from the fact that there are generically no degeneracies in the Floquet spectrum leading the off-diagonal matrix elements to time-average to zero. As Fig.~\ref{fig:imbalance} shows, $I_\infty$ quickly approaches 0 in the thermal phase even for $L = 13$, and the transition to a non-zero value in the MBL phase as a function of $\Gamma$ becomes steeper as $L$ is increased. The inset in Fig.~\ref{fig:imbalance} plots the same quantity for a ``canonical'' Hamiltonian model used in studies of MBL: 
\begin{equation}
H = \sum_i \mathbf{S}_i\cdot \mathbf{S}_{i+1}+ h_i S_i^z,
\label{eq:Ham}
\end{equation}
where $S_i^{\alpha = x,y,z} = \sigma_i^{\alpha}/2$ are spin 1/2 operators on site $i$ are the fields $h_i$ are drawn randomly and uniformly from $[-W, W]$.  The MBL phase transition in this model is at $W_c \gtrsim 3.5-4.5$\cite{PalHuse, Luitz15, Devakul15}.  As the inset shows, the trend with increasing $L$ is indeed towards $I_\infty \to 0$ in the thermal phase at small $W$. However, the conservation of energy and $S^z_{\rm tot}$ impedes thermalization\cite{LHD} and leads to comparatively much larger values of $I_\infty$ deep in the thermal phase for this Hamiltonian model as compared to our Floquet model, for similar system sizes. This supports our preconception that a Floquet model better highlights the contrasts between the MBL and thermal phases and can, thus, more sensitively probe the transition at small system sizes.

\begin{figure*}
\includegraphics[width=1.5\columnwidth]{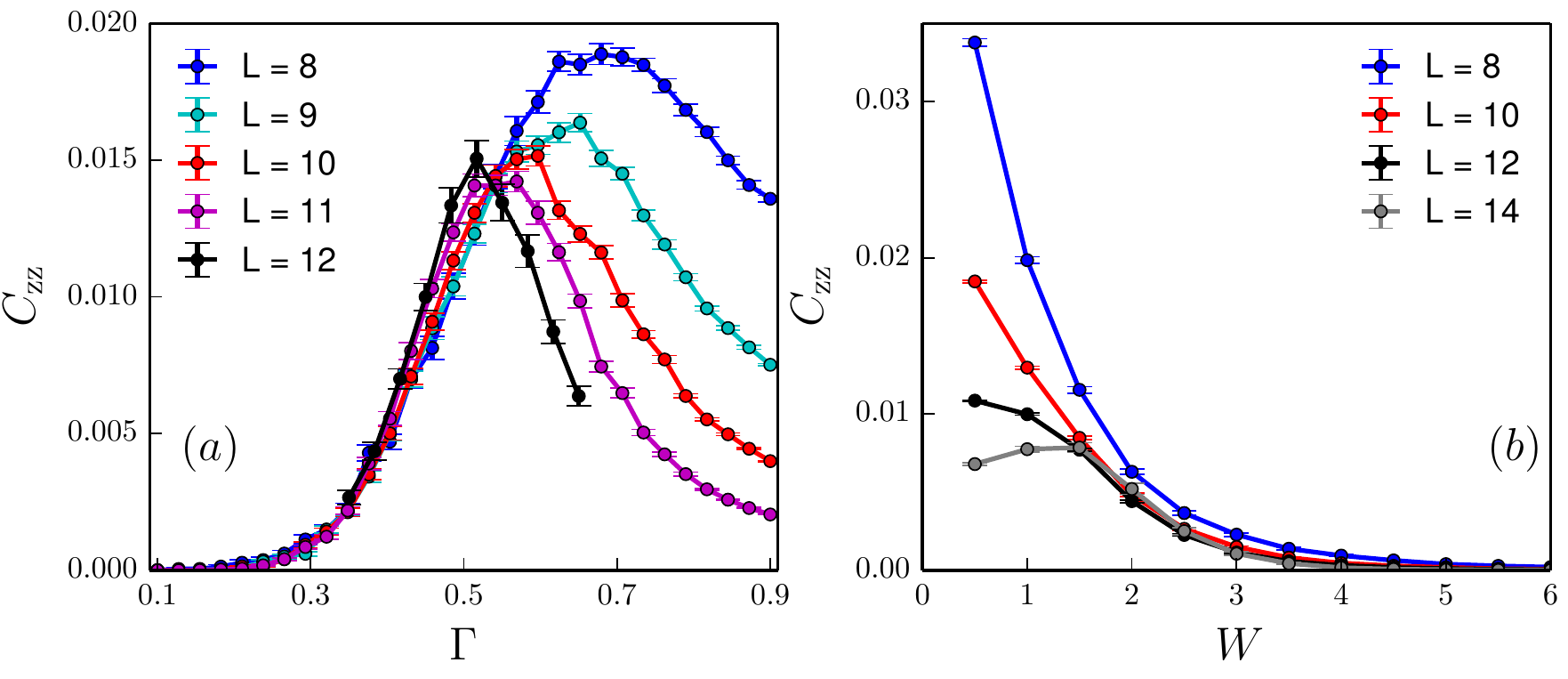}
\caption{(color online). The disorder and eigenstate averaged two-point connected correlation function $C_{zz}$ defined in \eqref{eq:czz} plotted for different system sizes in (a) the Floquet model \eqref{eq:model} and (b) the Hamiltonian model \eqref{eq:Ham}. This quantity shows a well-resolved peak near the MBL-to-thermal transition in the Floquet model even at small system sizes, while the peak is barely discernible in the Hamiltonian model.}
	\label{fig:czz}
\end{figure*}

\begin{figure*}
\centering
\includegraphics[width=1.5\columnwidth]{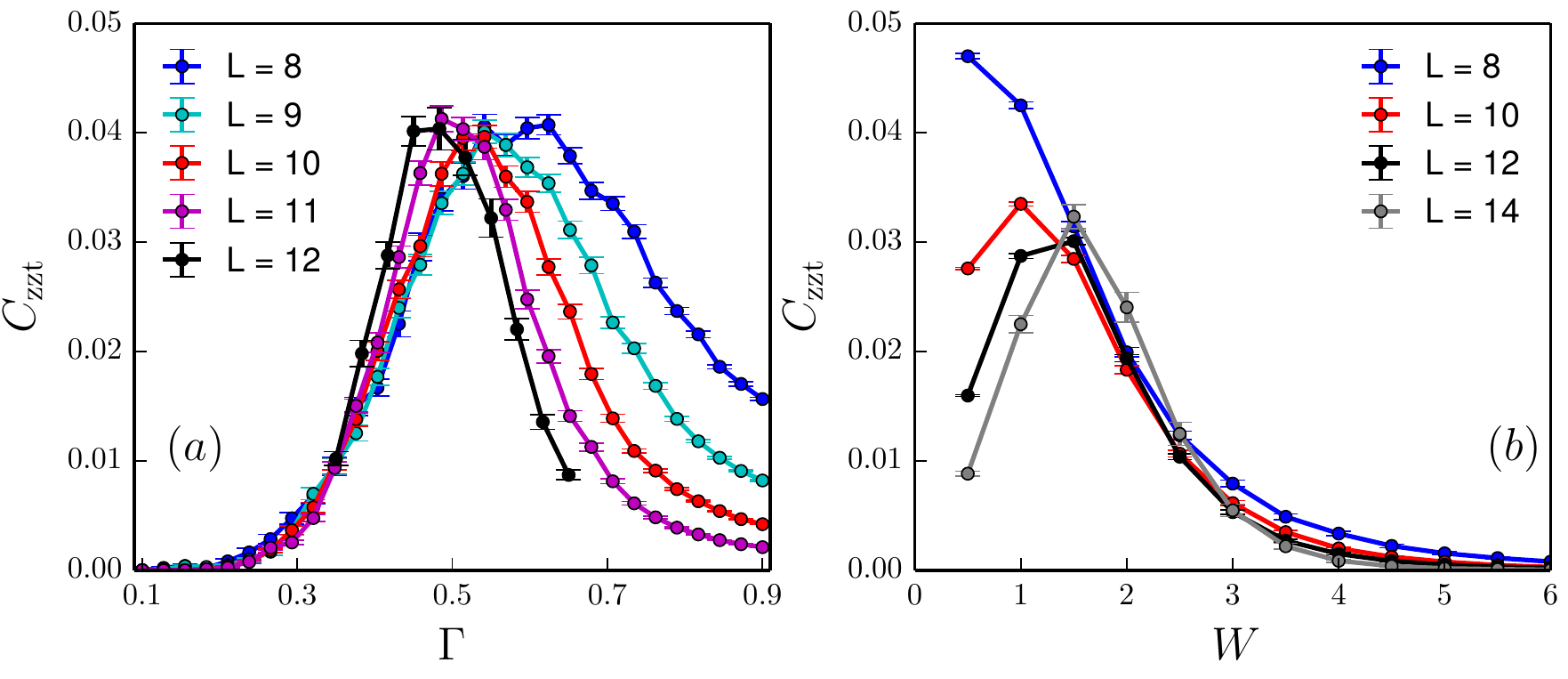}
\caption{(color online). The disorder and thermally averaged four-point spacetime connected correlation function $C_{zzt}$ defined in \eqref{eq:czzt} plotted for different system sizes in (a) the Floquet model \eqref{eq:model} and (b) the Hamiltonian model \eqref{eq:Ham}. This quantity shows a well-resolved peak at the MBL-to-thermal transition in the Floquet model even at small system sizes, while the peak is much weaker in the Hamiltonian model.}
	\label{fig:czzt}
\end{figure*}

\section{Correlation functions}
We now examine the behavior of two long-distance correlation functions in the thermal and localized phases and at the transition between them. Correlation functions have been a workhorse in the study of critical phenomena at ``conventional'' phase transtions, but have nevertheless been underused for probing the MBL transition. While correlators have certainly been discussed in the context of MBL\cite{PalHuse}, their utility for studying the transition has been limited by their sensitivity to conservation laws at small system sizes.

We start with the simplest spatially end-to-end connected correlation function\cite{PalHuse} averaged over eigenstates $\{\ket{n}\}$ and disorder:
\begin{equation}
\begin{aligned}
C_{zz} = \left[\frac{1}{2^L}\sum_n\left(\bra{n}\sigma_1^z\sigma_L^z\ket{n}-\bra{n}\sigma_1^z\ket{n}\bra{n}\sigma_L^z\ket{n}\right)^2\right].
\end{aligned}
\label{eq:czz}
\end{equation}
In the thermal phase at large $L$, this correlation function approaches the thermal equilibrium value appropriate to the ``infinite temperature'' ensemble which averages over all eigenstates equally. In the absence of any conservation laws , $\bra{n}\sigma_i^z\ket{n}$ and  $\bra{n}\sigma_i^z \sigma_j^z \ket{n}$ both decay exponentially with system size in a typical infinite temperature thermal eigenstate $\ket{n}$\cite{Deutsch, Srednicki}. 
In turn, $C_{zz}$  should decay exponentially with $L$ on a length scale set by some correlation length $\xi_{-}$ (the precise relationship between this $\xi_{-}$ and the longest diverging length scale associated with the MBL-to-thermal phase transition remains an open question). On the other hand, the eigenstates in the localized phase are not thermal and $\bra{n}\sigma_i^z\ket{n}$, $\bra{n}\sigma_i^z\sigma_j^z\ket{n}$ remain non-zero even for infinite $L$. Nevertheless, the typical correlation between two spins in the localized phase falls off exponentially with their separation, so that the connected correlator $C_{zz}$ still decays exponentially with $L$ on a length scale $\xi_{+}$ related to the localization length. Since $\xi_{-/+}$ increase as the transition is approached from both sides,  $C_{zz}$ also increases and shows a peak at the transition which sharpens with increasing $L$.  Fig.~\ref{fig:czz}(a) shows $C_{zz}$ plotted against $\Gamma$ for different $L$'s, and we note that the peak at the transition is quite sharp and well-resolved even at a modest system size of $L=12$. 

By contrast, $C_{zz}$ in the Hamiltonian model \eqref{eq:Ham} has a barely discernible peak even at $L=14$ and, as such, looks featureless through the transition at these sizes as shown in Fig.~\ref{fig:czz}(b). The reason is that the Hamiltonian model conserves $\sigma^z_{\rm tot}$ and in the (largest) sector with zero total $\sigma^z$, the conservation results in anticorrelations such that  $\bra{n}\sigma_1^z \sigma_L^z\ket{n} \sim 1/(L-1)$ in thermal eigenstates $\ket{n}$. The presence of these relatively large correlations that vanish only as a power of $L$ in the thermal phase masks the location of the peak, making it difficult to use this correlator as a diagnostic of the transition at the system sizes at our disposal. We note that the presence of energy conservation has a similar effect on the correlations, so simply breaking the conservation of $\sigma^z_{\rm tot}$ in a Hamiltonian system will not improve matters a whole lot.

Next, we introduce a four point unequal-time end-to-end connected correlation function which is averaged over infinite time. This quantity measures the correlations between end spins that are ``frozen'' in time: 
\begin{equation}
\begin{aligned}
C_{zzt} &= \Big[\lim_{T\to\infty}\frac{1}{T}\int_T^{2T} dt\Big( 
\langle\sigma^z_1\sigma^z_L\sigma^z_1(t)\sigma^z_L(t)\rangle \\ &-\langle\sigma^z_1\sigma^z_L\rangle\langle\sigma^z_1(t)\sigma^z_L(t)\rangle  -\langle\sigma^z_1\sigma^z_1(t)\rangle\langle\sigma^z_L\sigma^z_L(t)\rangle \\
&- \langle\sigma^z_1\sigma^z_L(t)\rangle\langle\sigma_L^z\sigma_1^z(t)\rangle \Big) \Big]
\end{aligned}
\label{eq:czzt}
\end{equation}
where $\langle \cdots \rangle$ is the infinite temperature ensemble average over all eigenstates, $[\cdots]$ refers to disorder averaging as before, and $\sigma_i^z(t)$ are operators in the Heisenberg picture at time $t$. Again, the off-diagonal matrix elements will time average to zero such that
\begin{widetext}
\begin{equation}
\begin{aligned}
C_{zzt} &=  \Big[ \frac{1}{2^L}\sum_n \bra{n}\sigma_1^z \sigma_L^z\ket{n}^2 
- \frac{1}{2^{2L}}\left(\sum_n\bra{n}\sigma_1^z\ket{n}^2 \right)\left(\sum_n\bra{n}\sigma_L^z\ket{n}^2\right)
- \left( \frac{1}{2^L}\sum_n\bra{n}\sigma_1^z\ket{n}\bra{n}\sigma_L^z\ket{n} \right)^2 \\
&- \frac{2}{2^{2L}}\sum_{n\neq m} \bra{n}\sigma_1^z\ket{m}^2\bra{n}\sigma_L^z\ket{m}^2 \Big]
\end{aligned}
\end{equation}
\end{widetext}
where we've made use of the reality of the eigenvectors $\{\ket{n}\}$ to drop absolute values. As before, $C_{zzt}$ decays exponentially in both the 
thermalizing and localized phases in models with no conservation laws, and shows a peak near the transition as a result of increased fluctuations and increased correlations in both space and time near the transition.  Fig~\ref{fig:czzt}(a) shows that $C_{zzt}$ has a very well resolved peak in the Floquet model even for $L=10$, and it is interesting to note that the height of this peak seems stable with system size for $L\geq 9$. 
Fig~\ref{fig:czzt}(b) plots $C_{zzt}$ in the Hamiltonian model \eqref{eq:Ham} with thermalization impeding conservation laws, and the resolution of the peaks at the transition is again quite poor when compared to the Floquet model.

\begin{figure}[t]
\includegraphics[width=\columnwidth]{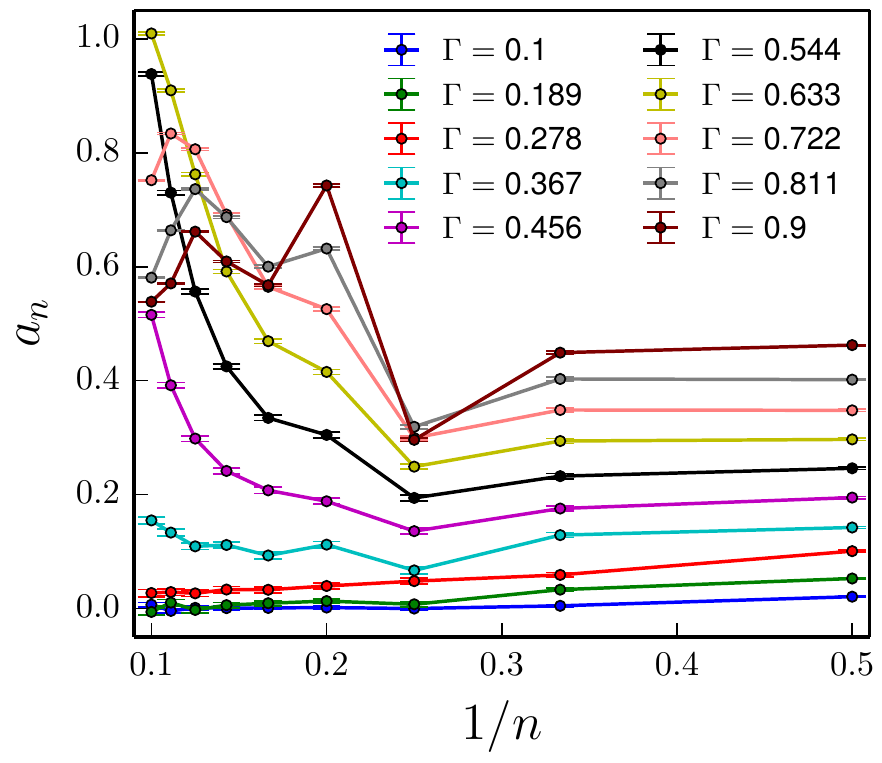}
\caption{(color online). The $n$th order numerical linked cluster contribution to the entanglement entropy averaged over 5000 disorder realizations. In the thermal phase, $ a_n $  approaches $ 1/2 $ as $ n\to\infty $, and in the localized phase $ a_n $  decreases exponentially to 0 for large enough $ n $.}
\label{fig:NLC}
\end{figure}

\section{Linked cluster expansion}
Recently Ref.~\onlinecite{Devakul15} used a numerical linked cluster expansion study of the entanglement entropy to obtain an estimate of an upper bound on the limit of stability of the MBL phase.
Here we apply the same calculation to our Floquet model at infinite temperature.
Following Ref.~\onlinecite{Devakul15}, we define
\begin{equation}
a_n = \sum_{c, |c|=n} \tilde{S}(c),\qquad \tilde{S}(c) = S(c) - \sum_{c^{'} \subset c} \tilde{S}(c^{'})
\end{equation}
at order $ n $, where $ S(c) $ is the bipartite von Neumann entanglement entropy across a pre-defined cut for a set of connected sites $ c $ across the cut, averaged over all eigenstates.  
Following Ref.~\onlinecite{Devakul15}, the disorder averaging over 5000 realizations is performed before the subgraph subtraction.
For thermal states following volume law entanglement, $ a_n $ is expected to saturate to a nonzero constant value at large $ n $\cite{Devakul15}.
In the localized phase, $ a_n $ is expected to decay exponentially to 0 when $ n $ is larger than some correlation length, so an extrapolation of $ a_{\infty}\geq 0 $ corresponds to a violation of the boundary law and thus a breakdown of localization \cite{Devakul15}.
The results for our model are shown in Fig.~\ref{fig:NLC}.
As $ \Gamma \to 1 $, $ a_n $ gradually settles down to about $ 1/2 $, demonstrating that our system is well thermalized in this region.
On the other end, as $ \Gamma\to 0 $, $ a_n $ steadily decreases, consistent with a transition from volume law entanglement to area law entanglement.
For small enough $ \Gamma $, we can see $ a_n $ suggests an extrapolation with $ a_{\infty} = 0 $, indicating the entry into localized regimes.
The upper bound for the transition in our Floquet model can thus can be crudely estimated as $ W\lesssim0.36 $.

\section{Conclusion}
In this paper, we propose and characterize a one-dimensional disordered Floquet model which can be used for studies of the phase transition between the thermal and MBL phases.
Because of the lack of conservation laws, the Floquet model is better thermalized compared to Hamiltonian models at finite sizes and thus better highlights the contrast between the MBL and thermal phases. We study several standard diagnostics of the MBL-to-thermal transition in this Floquet model, and show that these are well behaved even at relatively small system sizes. We also introduce a pair of long-distance correlation functions which show a sharp peak at the transition in the Floquet model, but are obscured by conservation laws in Hamiltonian models. In sum, we advocate for the use of Floquet models in studies of the MBL transition for mitigating the effects of finite-size, and also encourage more exploration into the behavior of physical correlation functions at the transition.   

\section{Acknowledgements}
We thank Trithep Devakul and S. L. Sondhi for helpful discussions.  D.A.H. was partially supported by the Addie and Harold Broitman Membership at I.A.S. V.K. was partially supported by NSF DMR-1311781 and by the Harvard Society of Fellows.

\bibliography{global}

\end{document}